# User equilibrium traffic assignment: *k* paths subtracting-adding algorithm

Miloš Nikolić[1]     Dušan Teodorović[1,*]

[1] University of Belgrade, Faculty of Transport and Traffic Engineering,
Vojvode Stepe 305, 11000 Belgrade, Serbia

**Abstract**

The traffic assignment problem is one of the most important transportation planning problems. The task faced by transportation planners, traffic engineers, and computer scientists is to generate high quality, approximate solutions of users equilibrium, that enable traffic scenario comparisons in a reasonable CPU time. We introduce the *k* **P**aths **S**ubtracting-**A**dding (*k*-PSA) algorithm to approximate the user equilibrium of the traffic assignment problem. The *k*-PSA algorithm consists of two alternating phases: (1) enlargement of the set of attractive paths; (2) subtracting-adding trips between generated attractive paths for each origin-destination pair of nodes. The proposed algorithm performs the two phases iteratively until the number of paths for each origin-destination pair is *k*. We tested the proposed algorithm on four benchmark transportation networks from the literature. The performed numerical tests show that the proposed approach generates, in short, computation times, solutions that are, on average, very close to the user equilibrium.

**Keywords:** Transportation Planning, Traffic Assignment Problem, User Equilibrium, Heuristic algorithm

**Introduction**

Numerous interventions in urban traffic networks (building new facilities, construction or rehabilitation work zones, congestion pricing cordons, etc.) modify the way that network users are distributed through the traffic network. Predicting future traffic flows in traffic networks represents a crucial component in transportation planning and traffic control. The main objective of traffic flows prediction is to develop user-friendly, precise, and reliable models, which can offer much information to planners and traffic engineers.

The urban traffic network consists of its set of nodes, set of links, link orientation, node connections, and link performance functions. These elements comprise transportation supply. The Origin-Destination matrix, and the link free-flow travel times that relate to the free-flow traffic conditions, describe transportation demand. The more vehicles pass through the link, the higher the level of traffic congestion and the higher the travel time. Link performance functions model the relationship between travel time and traffic volume on the links of a traffic network.

---
[*] Corresponding author
e-mail addresses: m.nikolic@sf.bg.ac.rs (Miloš Nikolić), duteodor@gmail.com (Dušan Teodorović)

The traffic assignment problem is, without a doubt, the fundamental transportation-planning problem. It could be defined in the following way: for the known transportation networks (characteristics of nodes and links) and transportation demand (Origin-Destination matrix), calculate link flows and link travel times. The traffic assignment could answer the following question: How are users distributed through the transportation network? Could users be distributed in many ways? The key computational challenge in traffic assignment procedures is to make a fast and precise estimate of traffic flows. In this way, the analyst can easily compare many network solutions. So far developed, traffic assignment procedures use many principles when distributing users. A mathematician Kohl (1841), and an economist Pigou (1918), generated the first ideas related to the traffic assignment problem. The user equilibrium (UE) and system optimal (SO) represent two essential traffic assignment models that have been developed to solve the traffic assignment problem. (Wardrop 1952). Wardrop's first principle is "The travel times on all used paths between an origin and a destination point are equal and less than those which would be experienced by a single vehicle on any unused path." Wardrop's second principle is related to the minimization of the total travel time in the network (system optimization formulation (SO)). SO formulation generates a smaller total number of hours traveled than the UE formulation. However, minimization of total travel time of all network users causes travel time increase of some drivers. These drivers could try to change the assigned path. In other words, an SO solution is hard to implement in real life.

When solving the traffic assignment problem, analysts assign the vehicles into the street network in such a way as to minimize a defined objective function. The objective function could be related to travel time, travel cost, air pollution, etc. The traffic assignment problem is an optimization problem that is difficult to solve, especially in large transportation networks. When solving UE or SO, the search space can be enormous. Consequently, many researchers use various heuristic approaches to find approximate solutions to the traffic assignment problem.

A most significant contribution of this paper is the development of the new algorithm for the static deterministic user-equilibrium traffic assignment problem. We propose the $k$-PSA algorithm. Within the $k$-PSA algorithm, we are iteratively doing two steps: first, we generate one new path between every origin-destination pair of nodes, and second, we determine traffic assignment using subtracting-adding procedure considering the generated paths. We also propose two new measures that enable quantitative performance evaluation of the $k$-PSA algorithm.

We present in this paper the numerical results (precision and speed) obtained for four analyzed well-known traffic networks (Sioux Falls, Barcelona, Chicago, and Chicago regional).
The rest of the paper is the following: Section 2 gives a short review of the traffic assignment problem; Section 3 contains the description of the $k$-PSA algorithm, while Section 4 describes the performance metrics of the $k$-PSA algorithm; Section 5 presents numerical results for five test networks, and Section 6 gives conclusions and recommendations for future research.

## 2. Traffic assignment problem: Short literature review

Campbell (1952) gave one of the first descriptions of the traffic assignment problem: "The estimated allocation of traffic to a proposed facility is commonly termed "traffic assignment." Beckmann et al. (1956) published their fundamental model and detailed analysis devoted to the traffic assignment problem. Their analysis included origin-destination matrix and flow-dependent link costs as inputs, and route and link flow as outputs. The current literature dedicated to the traffic assignment problems contains thousands of references (Knight 1924; Wardrop 1952; Beckmann et al. 1956; Jorgensen 1963; Dial 1971, 2006; LeBlanc et al. 1975; Florian and Nguyen 1976; Daganzo and Sheffi 1977; Florian 1977; Dafermos 1980; Friesz 1985; Sheffi 1985; Ben-Akiva and Lerman 1985; Teodorović and Kikuchi 1990; Larsson and Patriksson 1992; Patriksson 1994; Jayakrishnan et al. 1994; Henn 2000; Peeta and Ziliaskopoulos 2001; Bar-Gera 2002, 2006, 2010; Bar-Gera and Boyce 2003; Boyce et al. 2004; Yang and Huang 2004; Nie 2007, 2010; Florian et al. 2009, Teodorović and Janić 2016; Çolak et al. 2016; Lima et al. 2016).

The UE approach generates solutions from the drivers' behavior viewpoint. The basic UE assumption is that every network user tries to minimize his/her travel time (or some other factor that influences the user). Within the UE approach, every user is assigned to one of the best existing paths for its OD pair. The transportation network is in stable conditions when no user can reduce his/her travel time through the network by altering his/her path. These conditions are recognized as user equilibrium conditions (UE).

The majority of the developed route choice models that appeared in the literature in the last few decades are based on the assumption that people follow the shortest (minimum travel time) path. It is important to say that this assumption did not have significant empirical support in the past.

By following the GPS trajectories that explain the movement of personal cars, researchers, during the last decade, got a unique opportunity to better understand drivers' route choice behavior. The influential paper of Lima et al. (2016) is one of the best studies of the GPS trajectories. Lima et al. (2016) concluded, analyzing 92,419 anonym GPS trajectories during 18 months, that most drivers use a small number of paths for their daily trips and that many of them have a preferred route for everyday trips. Additionally, they discovered that "a significant fraction of drivers' routes is not optimal."

Merchant and Nemhauser (1978) introduced, and Peeta and Ziliaskopoulos (2001) analyzed and reviewed dynamic traffic assignment. Daganzo and Sheffi (1977) studied the stochastic aspect related to travel time perception. They formalized the concept of stochastic-user-equilibration, as an extension of Wardrop's user-equilibration criterion. Researchers have developed various microscopic simulation models related to traffic assignment during the last two decades. These modes incorporate traffic control devices, as well as interaction among vehicles.

## 3. The *k*-PSA algorithm

In the first step, most of the existing traffic assignment algorithms search for the set of paths that might be desirable to traffic network users. In the next step, analysts distribute trips to this set of paths. Finally, in the final step, analysts search for the convergence of the proposed procedure.

During the search for user equilibrium, we, all the time, expand the set of routes that might be desirable to traffic network users. We first generate the initial set of attractive paths. In the next step, we load the generated paths, and we discover the new shortest path. We include this path in a set of attractive paths (we increase the total number of attractive paths). We load the newly generated paths we again discover the new shortest path, etc. In other words, the path generation phase alternates with discovering the new shortest path phase.

We based our loading mechanism on the simple adding-subtracting scheme. To equalize the travel times among the alternative routes, we shift traffic flow between paths. We subtract part of the traffic flow from the longest path and add it to the shortest between the alternative paths.

Let us denote by *k* the maximum number of attractive paths that the analyst would like to generate for each O-D pair of nodes. By *n*, we denote the current number of generated paths for each O-D pair of nodes. The *k*-PSA algorithm, written more formally, reads:

Algorithm 1 - The *k*-PSA algorithm

Step 1: Treat travel time on all links as free flow travel time. Generate the shortest path for each O-D pair of nodes. Set that $n = 1$.
Step 2: Perform all or nothing traffic assignment.
Step 3: **while** $n < k$ **do**
Step 4: Generate new shortest path for each O-D pairs.
Step 5: Perform the new traffic assignment (load paths by using subtracting-adding procedure).
Step 6: $n = n + 1$
Step 7: **end while**

The analyst must specify in advance the total number of attractive paths *k* between each pair of nodes. We use Dijkstra's algorithm to determine the shortest paths. The first links and paths loads are the results of the all-or-nothing traffic assignment (step 2). Steps 4 and 5 are within the while loop. In step 4, we enlarge the set of attractive paths (we generate the new shortest path in the loaded network for each O-D pair of nodes). In step 5, we perform a new traffic assignment. We load paths by using the subtracting-adding procedure.

To explain the subtracting-adding procedure, let us introduce the following notation:

$N_n$ - number of subtracting-adding iterations in the case when there are *n* paths between each O-D pair of nodes

$\alpha_n$ - portion of demand that will be subtracted from one path and added the other path

$q_{rs}^{p}$ - part of demand from the source node *r* to destination node *s* that use path *p*
The following is the pseudo-code of the subtracting-adding procedure:

Algorithm 2 - Subtracting-adding procedure

Step 1: **for** i = 1 **to** $N_n$
Step 2: Determine travel times for all paths taking into account the last traffic assignment.
Step 3: **for** each O-D pair of nodes (*r*, *s*)
Step 4: Determine the path with the highest travel time ($p_1$) and the path with the lowest travel time ($p_2$). The new loads that will be assigned to paths read:
Step 5: $q_{rs}^{p_1,new} = q_{rs}^{p_1,old} - \alpha_n \cdot q_{rs}$
Step 6: $q_{rs}^{p_2,new} = q_{rs}^{p_2,old} + \alpha_n \cdot q_{rs}$
Step 7: **next**
Step 8: **next**

A general idea of the *k*-PSA algorithm is very simple. We send part of the demand from the path with the highest travel time to the path with the lowest travel time (Figure 1). In this way, we try to equalize travel times among paths as much as possible, i.e. we try to obtain the traffic assignment that is pretty close to the user equilibrium.

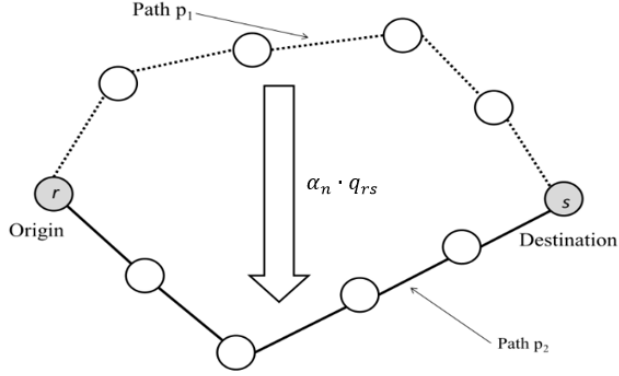

Figure 1. Sending part of demand from the path $p_1$ to the path $p_2$

From the pseudo-code of the subtracting-adding procedure, we see that the analyst should prescribe $N_n$ and $α_n$ parameters. Generally, from the experience that we gained from the experiments, $N_n$ could be the same for all values of $n$, but $α_{n+1}$ should be smaller than $α_n$.

## 4. Performance metrics of the *k*-PSA algorithm

The traffic assignment that represents user equilibrium is the final output of the proposed algorithm. According to Wardrop's first principle (1952), we do know that "the travel times on all used paths between an origin and destination point are equal, and less than those which would be experienced by a single vehicle on any unused path." Once, when we finish with the algorithm, we search again for the shortest paths in the loaded network for each O-D pair. One of the generated paths may be identical to the shortest path. In the cases of some O-D pairs, neither of the generated paths is identical to the shortest path. It is certainly necessary to determine how much the average travel time along generated paths differs from the travel time along the shortest path in the loaded network. The average travel time $\bar{t}_{rs}$ between node $r$ and node $s$ equals:

$$\bar{t}_{rs} = \frac{q_{rs}^{p_1} \cdot t_{rs}^{p_1} + \cdots + q_{rs}^{p_k} \cdot t_{ts}^{p_k}}{q_{rs}} \qquad (1)$$

For all O-D pairs, we calculate the relative difference (in percentage) $δ_{rs}$ between the average travel time along the generated paths, and the travel time along the shortest path in the loaded network:

$$δ_{rs} = \frac{\bar{t}_{rs} - t_{rs}}{t_{rs}} \cdot 100\% \quad \forall\, r, s \in OD \qquad (2)$$

where are:
$t_{rs}$ - travel time along the shortest path from the origin node $r$ to the destination node $s$
$OD$ - set of origin - destination pair of nodes

We produce histograms to visually illustrate the quality of the generated solutions, and we calculate the average error $E$ in percentages:

$$E = \frac{\sum_{rs \in OD} \delta_{rs}}{|OD|} \quad (3)$$

This error represents the first performance metric that we use.

We also made scatter diagrams and calculate the correlation coefficient to determine the strength of the relationship between the average travel time along generated paths and the travel time along the shortest path in the loaded network. The correlation coefficient $r$ value also represents the goodness of the discovered user equilibrium. If, for example, the correlation coefficient r equals 0.98, we denote discovered user equilibrium conditions as 0.98UE. If the correlation coefficient equals 0.99, we labeled found user equilibrium conditions as 0.99UE, etc. The correlation coefficient is the second performance metric that we use. The user equilibrium should have a correlation coefficient equal to 1.

The achieved CPU time also represents the performance metric that we use in this paper.

## 5. Numerical results

We present numerical results for four test transportation networks. The data related to these networks are available from www.bgu.ac.il/~bargera/tntp/. We implemented the $k$-PSA in the Java programming language. We performed all tests on desktop PS with the following performances: AMD Ryzen 7 3800X with 32 GB of RAM, operating system: Ubuntu 20.04.

Table 1 contains the data related to the set of problems (the number of nodes $|N|$, the number of links $|A|$, and the number of O-D pairs). The analyzed set of problems contains data related to Sioux Falls (USA), Barcelona (Spain), Chicago (USA), and Chicago regional (USA).

Table 1. Characteristics of test networks

| City | No. of nodes | No. of links | No. of OD pairs |
|---|---|---|---|
| Sioux Falls, USA | 24 | 76 | 528 |
| Barcelona, Spain | 1,020 | 2,522 | 7,922 |
| Chicago, USA | 933 | 2,950 | 93,135 |
| Chicago regional (USA) | 12,982 | 39,018 | 3,136,441 |
| Philadelphia (USA) | 13,389 | 40,003 | 1,150,722 |

Table 2 shows parameters of the subtracting-adding procedure that we used in this research.

Table 2. The values of the parameters used within the $k$-PSA algorithm

| $n$ | $N_n$ | $\alpha_n$ |
|---|---|---|
| 2 | 100 | 0.01 |
| 3 | 100 | 0.005 |
| 4 | 100 | 0.002 |
| 5 | 100 | 0.001 |

Table 3 shows the obtained relative deviations for the Sioux Falls benchmark network. We made three tests related to the Sioux Falls benchmark network. The number of paths $k$ for each O-D pair was respectively equal to 2, 3, and 4. Frequency column shows how many O-D pair has relative difference between average travel time and the shortest travel time within a specific percentage range. For example, in the experiment with four paths ($k = 4$), the 519 O-D pairs of nodes have an average deviation from 0 to 5 %. The relative frequency shows that 98.3 % of all O-D pairs of nodes belong to this deviation group. We obtained better results with the more attractive paths between pairs of nodes. Figure 2 gives the graphical representation of Sioux Falls' relative deviations.

Table 3. Sioux Falls network: Relative deviations between the average travel time along the generated paths, and the travel time along the shortest path in the loaded network

| Deviation | 2 paths ($k = 2$) | | 3 paths ($k = 3$) | | 4 paths ($k = 4$) | |
|---|---|---|---|---|---|---|
| | Frequency | Relative frequency | Frequency | Relative frequency | Frequency | Relative frequency |
| 0 % - 5 % | 470 | 0.890 | 506 | 0.958 | 519 | 0.983 |
| 5 % - 10 % | 27 | 0.051 | 12 | 0.023 | 4 | 0.008 |
| 10 % - 15 % | 10 | 0.019 | 8 | 0.015 | 3 | 0.006 |
| 15 % - 20 % | 7 | 0.013 | 2 | 0.004 | 1 | 0.002 |
| 20 % - 25 % | 4 | 0.008 | 0 | 0 | 1 | 0.002 |
| 25 % - 30 % | 2 | 0.004 | 0 | 0 | 0 | 0 |
| > 30 % | 8 | 0.015 | 0 | 0 | 0 | 0 |

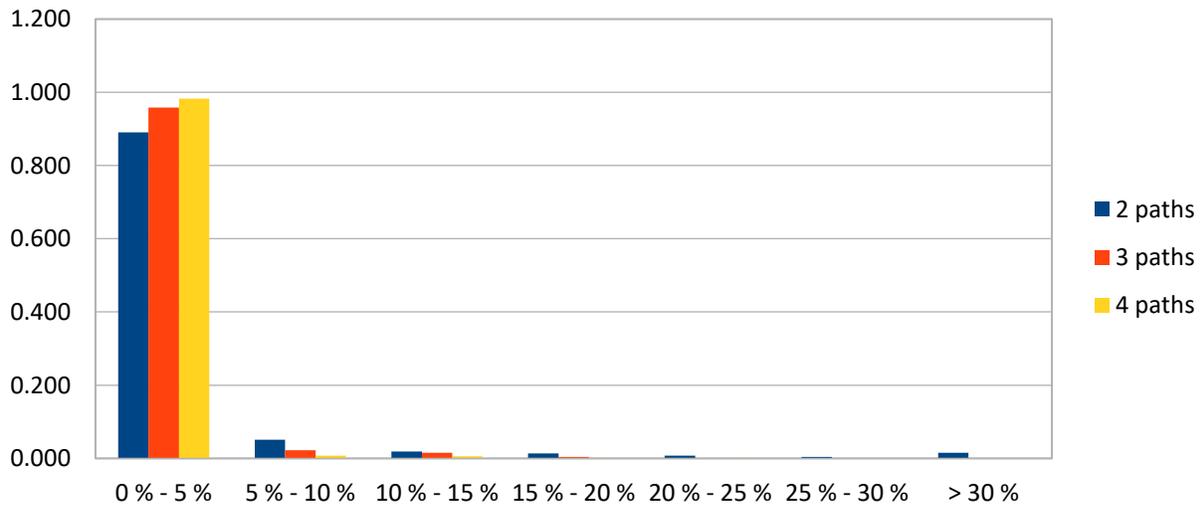

Figure 2. Relative deviations obtained for the Sioux Falls network

Table 4 shows the average relative deviation and CPU times obtained for the Sioux Falls network. We see from Table 4, that the $k$-PSA algorithm can find high quality solution for negligible CPU times.

Table 4. The average relative deviation and CPU times for the Sioux Falls network

| | 2 paths | 3 paths | 4 paths |
|---|---|---|---|
| Average relative deviation (%) | 2.268 | 0.583 | 0.326 |
| CPU times (ms) | 37 | 68 | 108 |

Table 5 and Figure 3 show the obtained frequencies and relative frequencies in the case of the Barcelona network. The number of paths $k$ for each O-D pair was also respectively equal to 2, 3, and 4. We perceive the results obtained in the case of the Barcelona network as very good. Almost all O-D pairs (99.8%) have a deviation in the interval from 0 to 5 %. Table 6 shows the average relative deviation. In all three cases, the average deviations are below 1 %.

Table 5. Barcelona network: Relative deviations between the average travel time along the generated paths, and the travel time along the shortest path in the loaded network

| Deviation | 2 paths | | 3 paths | | 4 paths | |
|---|---|---|---|---|---|---|
| | Frequency | Relative frequency | Frequency | Relative frequency | Frequency | Relative frequency |
| 0 % - 5 % | 7612 | 0.961 | 7882 | 0.995 | 7904 | 0.998 |
| 5 % - 10 % | 142 | 0.018 | 40 | 0.005 | 5 | 0.001 |
| 10 % - 15 % | 44 | 0.006 | 0 | 0 | 0 | 0 |
| 15 % - 20 % | 29 | 0.004 | 0 | 0 | 11 | 0.001 |
| 20 % - 25 % | 52 | 0.007 | 0 | 0 | 2 | 0.0003 |
| 25 % - 30 % | 18 | 0.002 | 0 | 0 | 0 | 0 |
| > 30 % | 25 | 0.003 | 0 | 0 | 0 | 0 |

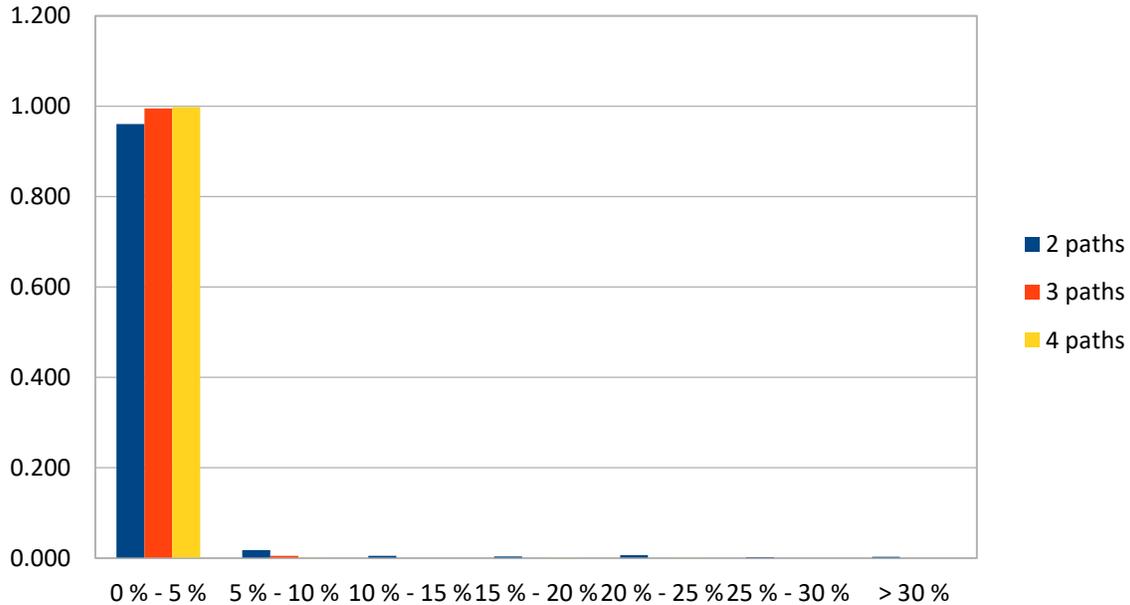

Figure 3. Relative deviations obtained for the Barcelona network

Table 6. The average relative deviation and CPU times for the Barcelona network

| | 2 paths | 3 paths | 4 paths |
|---|---|---|---|
| Average relative deviation (%) | 0.996 | 0.242 | 0.149 |
| CPU times (sec) | 1.027 | 1.957 | 3.978 |

Tables 7 and 8 and Figure 4 show the results obtained for the Chicago network. The quality of these results is similar to the quality of the results obtained for the Barcelona network. The algorithm needs slightly more CPU time for the Chicago network because it has 93,135 O-D pairs, while the Barcelona network has 7,922.

Table 7. Chicago network: Relative deviations between the average travel time along the generated paths, and the travel time along the shortest path in the loaded network

| Deviation | 2 paths | | 3 paths | | 4 paths | |
|---|---|---|---|---|---|---|
| | Frequency | Relative frequency | Frequency | Relative frequency | Frequency | Relative frequency |
| 0 % - 5 % | 88607 | 0.9514 | 92501 | 0.9932 | 92680 | 0.9951 |
| 5 % - 10 % | 3527 | 0.0379 | 502 | 0.0054 | 187 | 0.0020 |
| 10 % - 15 % | 847 | 0.0091 | 92 | 0.0010 | 94 | 0.0010 |
| 15 % - 20 % | 130 | 0.0014 | 40 | 0.0004 | 45 | 0.0005 |
| 20 % - 25 % | 22 | 0.0002 | 0 | 0 | 78 | 0.0008 |
| 25 % - 30 % | 2 | 0.00002 | 0 | 0 | 13 | 0.0001 |
| > 30 % | 0 | 0.000 | 0 | 0 | 38 | 0.0004 |

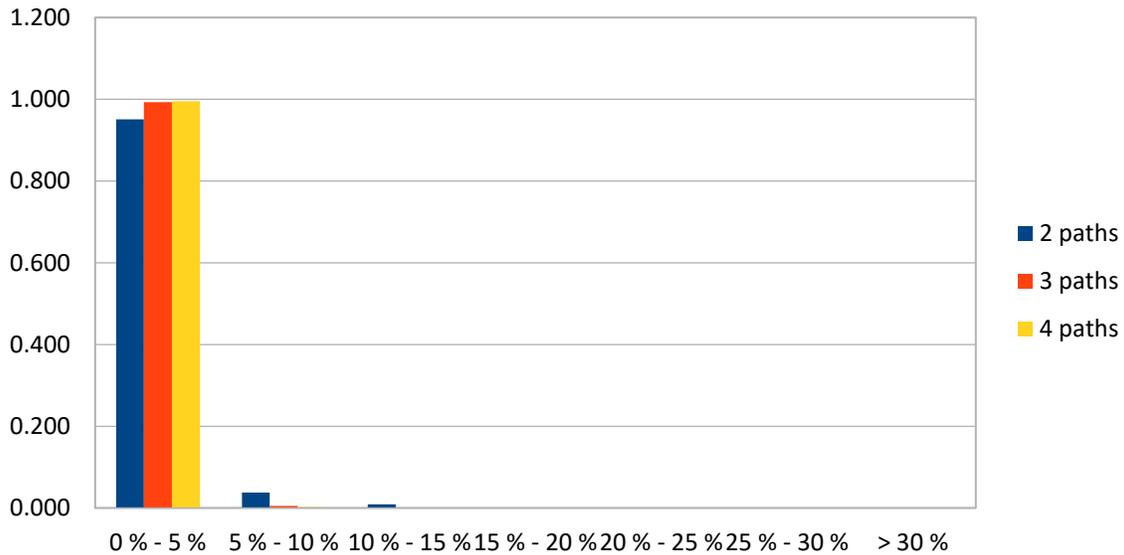

Figure 4. Relative deviations obtained for the Chicago network

Table 8. The average relative deviation and CPU times for the Chicago network

| | 2 paths | 3 paths | 4 paths |
|---|---|---|---|
| Average relative deviation (%) | 0.792 | 0.269 | 0.157 |
| CPU times (sec) | 4.548 | 10.703 | 24.829 |

We also tested our algorithm in the case of a huge benchmark network, Chicago regional. This network has over 3 million O-D pairs. The number of paths $k$ for each O-D pair was respectively equal to 2, 3, 4, and 5. Tables 9 and 10 and Figure 5 show obtained results. From the obtained results, we see that the total number of paths highly influences the quality of the solution. The average relative deviation for the case when $k = 2$ is almost 25 %. For $k = 3$ the deviation is 5.4 %, for $k = 4$ the deviation is 2.1 % and for $k = 5$, the deviation is only 1.25 %. These results clearly show that for larger networks analysts should consider more paths than in cases of smaller traffic networks.

Table 9. Chicago regional network: Relative deviations between the average travel time along the generated paths, and the travel time along the shortest path in the loaded network

| Deviation | 2 paths | | 3 paths | | 4 paths | | 5 paths | |
|---|---|---|---|---|---|---|---|---|
| | Frequency | Relative frequency | Frequency | Relative frequency | Frequency | Relative frequency | Frequency | Relative frequency |
| (0 - 5) % | 573366 | 0.250 | 1411703 | 0.615 | 2008065 | 0.875 | 2204091 | 0.960 |
| (5 - 10) % | 289982 | 0.126 | 467448 | 0.204 | 235237 | 0.102 | 85972 | 0.037 |
| (10 - 15) % | 257711 | 0.112 | 227414 | 0.099 | 45015 | 0.020 | 5409 | 0.002 |
| (15 - 20) % | 201177 | 0.088 | 100043 | 0.044 | 6920 | 0.003 | 224 | 0.000 |
| (20 - 25) % | 163626 | 0.071 | 47566 | 0.021 | 803 | 0.0003 | 390 | 0.0002 |
| (25 - 30) % | 135386 | 0.059 | 23871 | 0.010 | 150 | 0.00007 | 48 | 0.00002 |
| > 30 % | 674979 | 0.294 | 18182 | 0.008 | 37 | 0.00002 | 93 | 0.00004 |

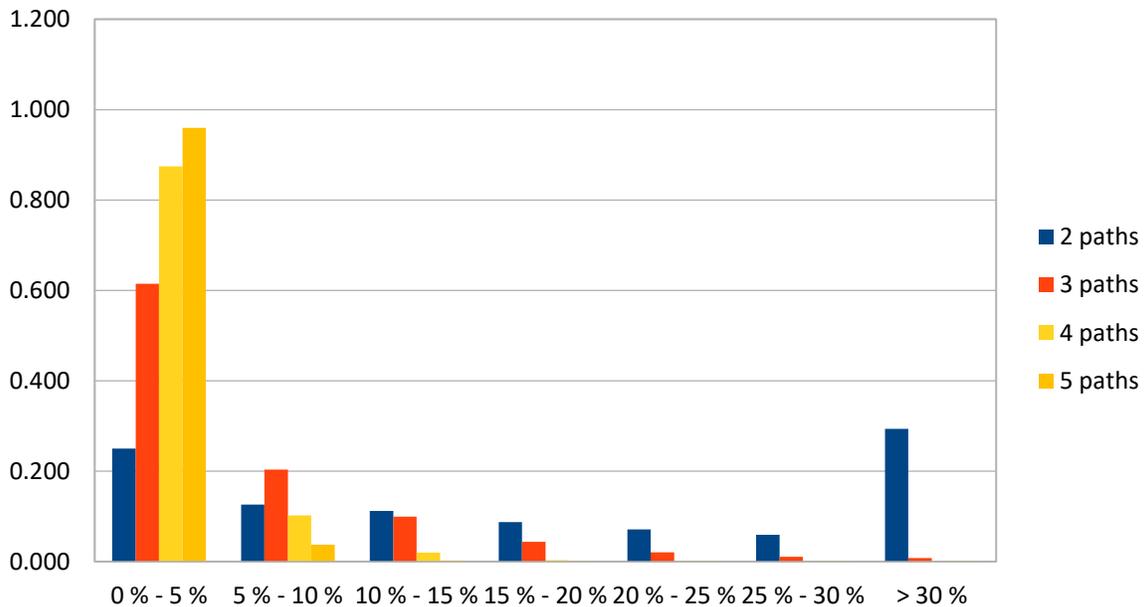

Figure 5. Relative deviations obtained for the Chicago regional network

Table 10. The average relative deviation and CPU times for the Chicago regional network

|  | 2 paths | 3 paths | 4 paths | 5 path |
|---|---|---|---|---|
| Average relative deviation (%) | 24.615 | 5.373 | 2.091 | 1.25 |
| CPU times (min) | 39.83 | 69.16 | 99.1 | 136.97 |

A correlation coefficient can also measure the quality of solutions obtained by *the k*-PSA algorithm. We created scatter diagrams for the Sioux Falls network. In the next step, we calculated the correlation coefficient to determine the strengths between the average travel time along generated paths and the travel time of the shortest path in the loaded network. Figures 6, 7, and 8 give examples of scatter diagrams. In these diagrams, the average travel time is on *the x*-axis, and the shortest travel time on *the y*-axis.

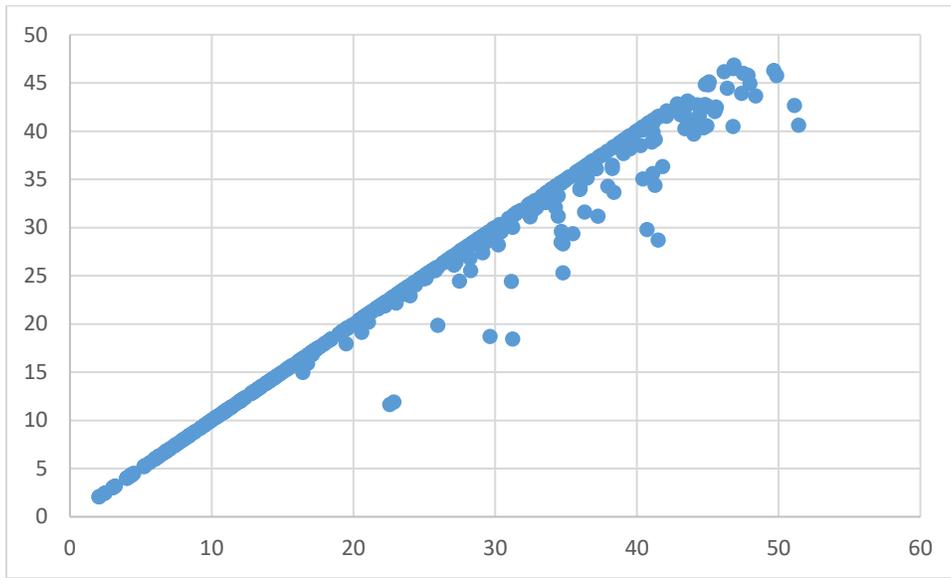

Figure 6. Correlation for the Sioux Falls network in the case of $k = 2$ ($R = 0.98941$)

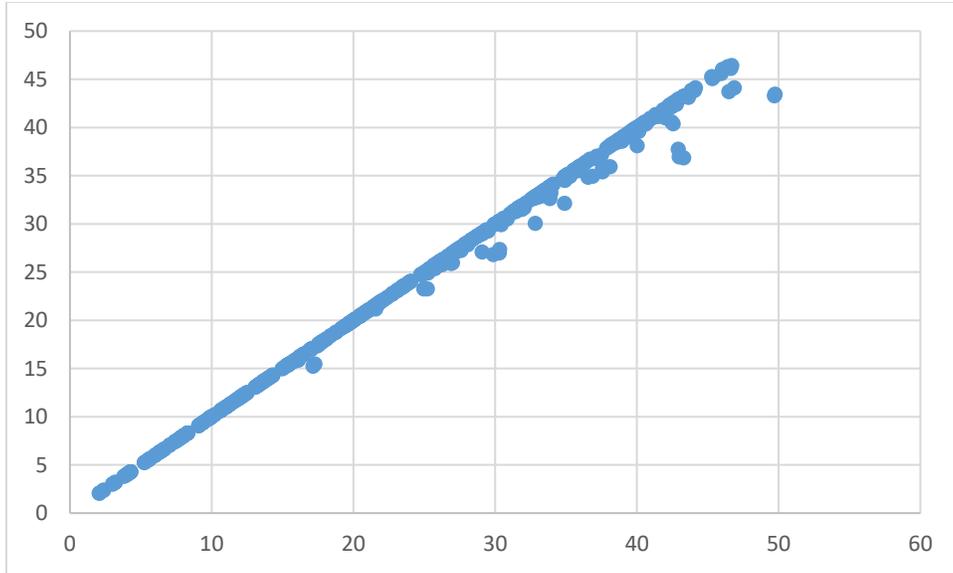

Figure 7. Correlation for the Sioux Falls network in the case of $k = 3$ ($R = 0.99814$)

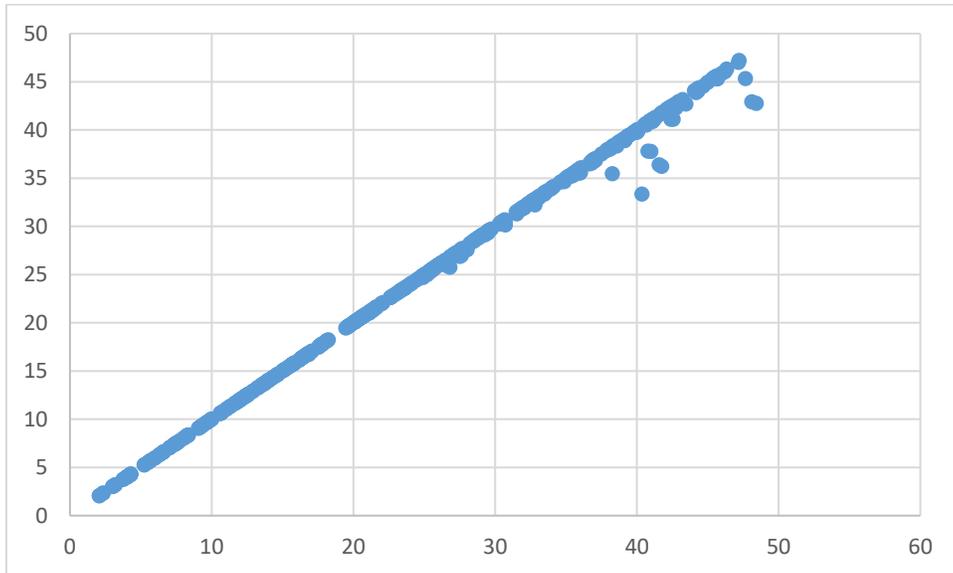

Figure 8. Correlation for the Sioux Falls network in the case of $k = 4$ ($R = 0.99868$)

In the example of Sioux Falls, we see that with the increase in the number of paths, the quality of the solution generated by the proposed $k$-PSA algorithm also increases. The correlation coefficient $r$ measure this quality. The cases of other analyzed transportation networks show that an increase in the number of used paths increases precision. Our results, obtained after several numerical experiments, are in agreement with the results obtained by Lima et al. (2016) after monitoring the behavior of thousands of drivers in several cities. Lima et al. (2016) experimentally confirmed that most drivers use a small number of paths for their daily trips. In addition, the proposed $k$-PSA algorithm generates traffic assignments very similar to user equilibrium already in cases when there are only a few paths between each O-D pair.

# 6. Conclusions

In this paper, we propose a novel algorithm to approximate the user equilibrium of the static traffic assignment problem. We call this algorithm - the $k$-PSA algorithm. The developed algorithm is capable of generating high quality, approximate solutions of users' equilibrium. On average, the generated solutions are very close to the user equilibrium. The $k$-PSA algorithm enables fast traffic scenario comparisons. The paper introduced novel performance metrics. We measure the average relative deviation between the average travel time along generated paths and the travel time along the shortest path in the loaded network. We compute the correlation coefficient to determine the strength of the relationship between the average travel time along used paths and the travel time along the shortest path in the loaded network. The correlation coefficient $r$ value also denotes the goodness of the discovered user equilibrium.

We applied the proposed algorithm on four transportation networks (one small size, two medium sizes, and one large). The obtained results clearly show that solutions are very close to the used equilibrium. The algorithm can find solutions for small and medium-size networks within negligible CPU times.

The measures $E$ and $r$ that we proposed to enable easy measurement of the quality of the obtained solution. They also permit the comparison of solutions for various networks. They measure the proximity of the obtained solution to the user equilibrium.

There is still a lot of space for modifications or improvements to the proposed approach. In future research, the proposed $k$-PSA algorithm may be part of the more complex algorithms for solving various combinatorial optimization problems on medium-sized traffic networks.

**Acknowledgement**

The Ministry of Education, Science and Technological Development of the Republic of Serbia, through the University of Belgrade, Faculty of Transport and Traffic Engineering, have supported this research.